\documentclass[floatfix,aps,prl,preprint,superscriptaddress]{revtex4-2}
\usepackage{upgreek}
\usepackage{graphicx}
\usepackage[]{epsfig}
\usepackage{times,amsmath,amssymb}

\usepackage{color}
\usepackage[colorlinks = true,
            linkcolor = blue,
            urlcolor  = blue,
            citecolor = blue,
            anchorcolor = blue]{hyperref}

\begin{document}

\title{Probing electron trapping by current collapse in GaN/AlGaN FETs utilizing quantum transport characteristics}

\author{Takaya Abe}
\affiliation{Research Institute of Electrical Communication, Tohoku University, 2-1-1 Katahira, Aoba-ku, Sendai 980-8577, Japan}
\affiliation{Department of Electronic Engineering, Graduate School of Engineering, Tohoku University, Aoba 6-6-05, Aramaki, Aoba-Ku, Sendai 980-8579, Japan}

\author{Motoya Shinozaki}
\email[]{motoya.shinozaki.c1@tohoku.ac.jp}
\affiliation{WPI Advanced Institute for Materials Research, Tohoku University, 2-1-1 Katahira, Aoba-ku, Sendai 980-8577, Japan}

\author{Kazuma Matsumura}
\affiliation{Research Institute of Electrical Communication, Tohoku University, 2-1-1 Katahira, Aoba-ku, Sendai 980-8577, Japan}
\affiliation{Department of Electronic Engineering, Graduate School of Engineering, Tohoku University, Aoba 6-6-05, Aramaki, Aoba-Ku, Sendai 980-8579, Japan}

\author{Takumi Aizawa}
\affiliation{Research Institute of Electrical Communication, Tohoku University, 2-1-1 Katahira, Aoba-ku, Sendai 980-8577, Japan}
\affiliation{Department of Electronic Engineering, Graduate School of Engineering, Tohoku University, Aoba 6-6-05, Aramaki, Aoba-Ku, Sendai 980-8579, Japan}

\author{Takeshi Kumasaka}
\affiliation{Research Institute of Electrical Communication, Tohoku University, 2-1-1 Katahira, Aoba-ku, Sendai 980-8577, Japan}

\author{Norikazu Ito}
\affiliation{ROHM Co., Ltd, 21 Saiinnmizosakicho, Ukyo-ku, Kyoto, Kyoto 615-8585, Japan}

\author{Taketoshi Tanaka}
\affiliation{ROHM Co., Ltd, 21 Saiinnmizosakicho, Ukyo-ku, Kyoto, Kyoto 615-8585, Japan}

\author{Ken Nakahara}
\affiliation{ROHM Co., Ltd, 21 Saiinnmizosakicho, Ukyo-ku, Kyoto, Kyoto 615-8585, Japan}

\author{Tomohiro Otsuka}
\email[]{tomohiro.otsuka@tohoku.ac.jp}
\affiliation{WPI Advanced Institute for Materials Research, Tohoku University, 2-1-1 Katahira, Aoba-ku, Sendai 980-8577, Japan}
\affiliation{Research Institute of Electrical Communication, Tohoku University, 2-1-1 Katahira, Aoba-ku, Sendai 980-8577, Japan}
\affiliation{Department of Electronic Engineering, Graduate School of Engineering, Tohoku University, Aoba 6-6-05, Aramaki, Aoba-Ku, Sendai 980-8579, Japan}
\affiliation{Center for Science and Innovation in Spintronics, Tohoku University, 2-1-1 Katahira, Aoba-ku, Sendai 980-8577, Japan}
\affiliation{Center for Emergent Matter Science, RIKEN, 2-1 Hirosawa, Wako, Saitama 351-0198, Japan}


\begin{abstract}
GaN is expected to be a key material for next-generation electronics due to its interesting properties. 
However, the current collapse poses a challenge to the application of GaN FETs to electronic devices. 
In this study, we investigate the formation of quantum dots in GaN FETs under the current collapse. 
By comparing the Coulomb diamond between standard measurements and those under current collapse, we find that the gate capacitance is significantly decreased by the current collapse.
This suggests that the current collapse changes the distribution of trapped electrons at the device surface, which is reported in the previous study by operando X-ray spectroscopy. 
Also, we show external control of quantum dot formation, previously challenging in an FET structure, by using current collapse.
\end{abstract}

\maketitle

Gallium nitride (GaN), a wide bandgap semiconductor featuring direct transitions, is renowned for forming high-mobility two-dimensional electron gases (2DEGs) at the GaN/AlGaN heterostructure interface~\cite{ambacher1999two, manfra2004electron, thillosen2006weak, shchepetilnikov2018electron}. 
This property positions GaN as a pivotal material for next-generation electronics, including power electronics and 5G technology applications.
In the high power operation of GaN field-effect transistors (FETs), a challenge arises from a phenomenon known as current collapse~\cite{jones2016review, binari2002trapping}. 
Current collapse induces an increase in on-resistance and fluctuations of the output current by applying high electric fields. 
This phenomenon critically hampers the device's ability to perform as expected. 
It is suggested that the surface states of the device and the interface states between the gate insulator and AlGaN significantly contribute to this issue.
Current collapse poses a key issue for the application of GaN FETs into electronic devices~\cite{jones2016review}. 
Its mechanism has been investigated through various methods, including transport measurements~\cite{bisi2013deep, tokuda2016dlts, yang2019characterization, yang2020experimental, pan2021identifying, pan2022characterization} and operando X-ray spectroscopy~\cite{omika2018operation, omika2020dynamics}.
To gain a more microscopic view, operando X-ray spectroscopy is a powerful technique for probing the electron state with high temporal and spatial resolution~\cite{fukidome2014pinpoint}, which local information is difficult to access by the typical electron transport measurement.
Thanks to various approaches, the mechanism of the current collapse is becoming increasingly clear.

From the perspective of quantum devices~\cite{Tarucha1996, kouwenhoven2001few}, GaN is attractive because of its wide and direct bandgap.
These characteristics can provide high-temperature operation and coupling with light~\cite{luo2024room}.
Previously, quantum transport is observed in GaN devices with a gate voltage defined~\cite{chou2005high, chou2006single}, nano-sturcutures~\cite{ristic2005columnar, nakaoka2007coulomb, Songmuang2010}, or a simple FET structures~\cite{Otsuka2020, matsumura2023channel, fujiwara2023wide}.
In particular, orbital excited states is also observed in the FET structure with fine gate electrode~\cite{matsumura2023channel}.
This result opens the fascinating application of the GaN quantum devices, such as a quantum bit operation.
Quantum dots in the simple FET structure are formed by disordered potential fluctuation induced by impurities and defects~\cite{Otsuka2020, matsumura2023channel}.
This mechanism suggests that surface electron trapping caused by the current collapse also affects the characteristics of the quantum dots.
By utilizing this sensitivity, it is possible to probe the local distribution of surface electrons.
In this study, we shed light on the microscopic perspective of the distribution of surface electrons by analyzing electron transport through GaN quantum dots under the current collapse.

\begin{figure}
\begin{center}
  \includegraphics{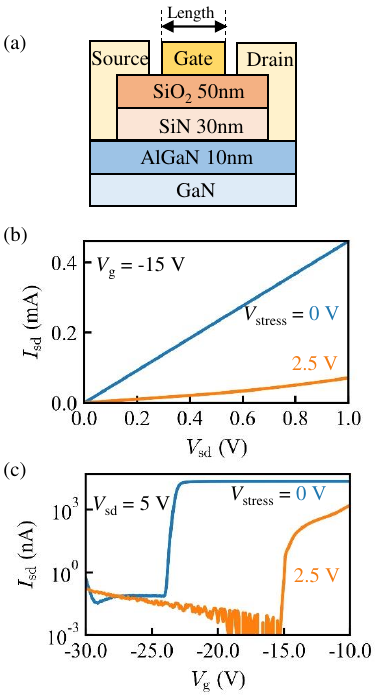}
  \caption{(a) Schematic of the layer structure of the device.
  (b) $I_\mathrm{sd}$-$V_\mathrm{sd}$ and (c) $I_\mathrm{sd}$-$V_\mathrm{g}$ characteristics with and without the stress voltage at 2.3 K.}
  \label{fig1}
\end{center}
\end{figure}

Fig.~\ref{fig1}(a) shows a layer structure of our device. 
GaN and AlGaN layers are deposited on the Si substrate by chemical vapor deposition. 
The 2DEG is formed at the interface between the GaN and AlGaN layers. 
The typical electron density and mobility of our layer structure are $6.7\times10^{12}\;\mathrm{cm}^{-2}$ and $1670 \;\mathrm{cm}^{2}\mathrm{V}^{-1}\mathrm{s}^{-1}$, respectively. 
The source and drain electrodes are constructed using Ti/Al metals. 
The devices are fabricated with gate lengths of 0.6, 0.8, and 1.4 $\upmu$m.
In order to induce the current collapse, a stress voltage $V_\mathrm{stress}$ is applied to the source electrode in the pinch-off state at 300 K, which state is maintained for 1 hour.
Subsequently, devices are cooled from 300 K to 2.3 K using a helium decompression refrigerator, spending approximately 2 hours keeping the $V_\mathrm{stress}$.

After cooling, the source-drain current $I_\mathrm{sd}$ is measured as a function of the source-drain voltage $V_\mathrm{sd}$, and a gate voltage $V_\mathrm{g}$ near the pinch-off state.
The gate length of the measured device is 0.6 $\upmu$m.
Figure~\ref{fig1}(b) shows the $I_\mathrm{sd}$-$V_\mathrm{sd}$ characteristics at 2.3 K. 
The blue line indicates the result without applying a stress voltage, while the orange line with $V_\mathrm{stress} = 2.5 \, \mathrm{V}$. 
During the measurement, $V_\mathrm{g}$ is set to the 'on' state at $V_\mathrm{g} = -15 \, \mathrm{V}$. 
When the stress voltage is applied, compared to the standard measurement, $I_\mathrm{sd}$ is significantly reduced, and an increase in the on-resistance is observed. 
The current collapse is caused by cooling the sample with the application of stress voltage.
Figure~\ref{fig1}(c) also shows the $V_\mathrm{g}$ dependence of $I_\mathrm{sd}$ at 2.3 K.
The blue line represents the results from the standard measurement, while the orange line indicates the current collapse condition.
Here, the $V_\mathrm{sd}$ is set to 5 V. 
The application of stress voltage results in a shift of the pinch-off voltage towards the positive side, which is approximately +10 V.
The shift in pinch-off due to the thermal cooling from 300 K to 2.3 K is typically around 1.5 V in our devices.
Therefore, this large change of shift voltage is due to electrons trapped at the AlGaN/SiN interface and/or defect levels in the insulator.

\begin{figure}
\begin{center}
  \includegraphics{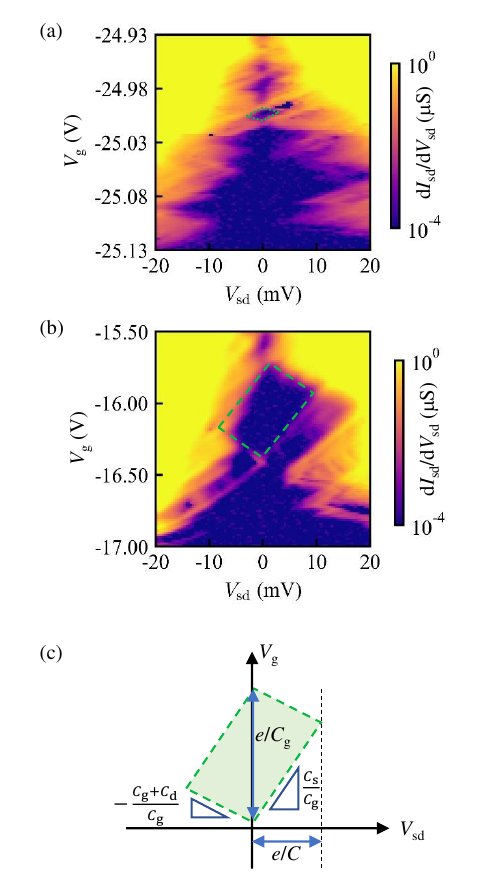}
  \caption{(a) Color map of the $\mathrm{d} I_\mathrm{sd}/\mathrm{d} V_\mathrm{sd}$ as a function of $V_\mathrm{sd}$ and $I_\mathrm{g}$ in a standard measurement and (b) under the current collapse. The dashed green area corresponds to the analyzed diamond. 
  (c) Schematic of the Coulomb diamond.}
  \label{fig2}
\end{center}
\end{figure}

Figure~\ref{fig2}(a) shows the results of the FET characteristics near the pinch-off voltage for the device with a gate length of 0.6 $\upmu$m, and the result measured under the current collapse condition ($V_\mathrm{stress}$ = 2.5 V) is also displayed in Fig~\ref{fig2}(b).
The differential conductance $\mathrm{d}I_\mathrm{sd}/\mathrm{d}V_\mathrm{sd}$ is plotted on a logarithmic scale against $V_\mathrm{sd}$ and $V_\mathrm{g}$.
In both cases, the $\mathrm{d}I_\mathrm{sd}/\mathrm{d}V_\mathrm{sd}$ is suppressed, displaying unique shapes as shown in the figures.
These are Coulomb diamonds due to the formation of quantum dots in the conduction channel~\cite{Otsuka2020, matsumura2023channel}.
The shape of the Coulomb diamond under the current collapse, especially in the $V_\mathrm{g}$ axis direction, is about five times larger than the standard measurement.
Table~\ref{table1} shows the electrostatic capacitances between the quantum dot and each electrode, the total electrostatic capacitance $C$, and the coefficient $\alpha$ that converts gate voltage to electrostatic energy called lever arm~\cite{ono2013pseudosymmetric}. 
Each capacitance is determined from the shape of the Coulomb diamond by assuming the capacitive model~\cite{kouwenhoven2001few, nuryadi2003ambipolar, song2015gate, muto2024visual}, as shown in Fig.~\ref{fig2}(c).
The electrostatic capacitance $C_\mathrm{g}$ between the quantum dot and the gate electrode significantly decreases under current collapse.
$\alpha$ also decreases by two orders of magnitude, indicating a reduced contribution of the $C_\mathrm{g}$ to the total electrostatic capacitance $C$. 
Furthermore, $C_\mathrm{s}$ and $C_\mathrm{d}$, the electrostatic capacitances between the quantum dot and the source and drain electrodes, are close in the standard measurement, while $C_\mathrm{s}$ is about three times larger than $C_\mathrm{d}$ under the current collapse.

\begin{table}
  \caption{Summary of electrostatic capacitances. Units of each capacitance are aF.}
  \label{table1}
  \vspace{2mm}
  \centering
  \begin{tabular}{cccccc}
    & $C_\mathrm{g}$  & $C_\mathrm{s}$  & $C_\mathrm{d}$  & $C$  & $\alpha = C_\mathrm{g}/C$\\
  \hline\hline
  Standard & 8.3 & 25.8 & 23.8 & 56.7 & 0.15 \\
  \hline
  Collapse & 0.2 & 10.3 & 3.3 & 13.8 & 0.0016 \\
  \hline
  \end{tabular}
\end{table}

\begin{figure}
\begin{center}
  \includegraphics{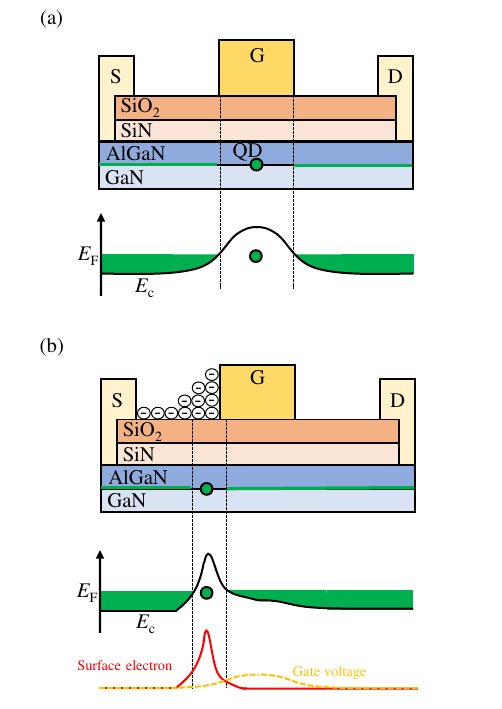}
  \caption{Schematic of the surface electron trapping and an energy band structure (a) in the case of the standard measurement and (b) the current collapse.}
  \label{fig3}
\end{center}
\end{figure}

The change in electrostatic capacitance can be qualitatively understood and is consistent with the insights of previous studies using operando X-ray spectroscopy~\cite{omika2018operation, omika2020dynamics}.
Near the pinch-off voltage under the standard measurement, the spatial profile of the conduction band's bottom energy level $E_\mathrm{c}$ is as shown in Fig.~\ref{fig3}(a).
The quantum dot is formed directly below the gate electrode and is almost symmetrically connected to both the source and drain electrodes.
When the stress voltage is applied to the source electrode, electrons are trapped in surface levels towards the source from the gate electrode, as illustrated in Fig.~\ref{fig3}(b).
This electron trapping causes the change in the $E_\mathrm{c}$ profile.
$E_\mathrm{c}$ rises steeply at the edge of the gate electrode and decreases gradually towards the source electrode.
Therefore, near the pinch-off voltage under current collapse, it is expected that the formation of the quantum dot is not directly below the gate electrode but near the edge of the gate electrode where electron trapping occurs.
Consequently, the coupling between the gate electrode and the quantum dot becomes smaller under current collapse, resulting in the decrease in $C_\mathrm{g}$.
Additionally, due to the superposition of the surface electron potential and the gate voltage set near the pinch-off region, an asymmetric coupling of the quantum dot and electrodes occurs, leading to the imbalance of $C_\mathrm{d}$ and $C_\mathrm{s}$.
These results can serve as a probing tool to investigate the distribution of surface electron trapping, which was previously challenging to access with simple transport measurements.

\begin{figure}
\begin{center}
  \includegraphics{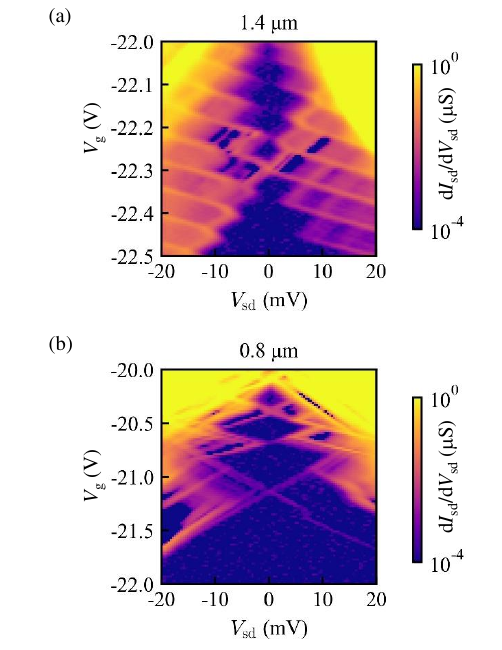}
  \caption{The gate length dependence of the Coloumb diamond. Each length is (a) 1.4 and (b) 0.8 $\upmu$m, respectively. 
  }
  \label{fig4}
\end{center}
\end{figure}

Finally, we measure the gate length dependence of the Coulomb diamond under the current collapse.
The stress voltage is set to be 5 V in all devices.
Clearly seen in Fig.~\ref{fig4}, both devices show a completely closed Coulomb diamond.
In the case of the standard condition, such a closed diamond is observed in the device with gate length below 0.2 $\upmu$m~\cite{matsumura2023channel}.
This result suggests that the current collapse makes an effective short gate length structure in the vicinity of the gate electrode edge, as shown in Fig.~\ref{fig3}(b).
This built-in short gate forms a locally confined structure regardless of the device's gate length, resulting in closed diamonds even in devices with longer gate lengths.

In this study, we investigate the formation of quantum dots under the current collapse in GaN FETs.
The devices are cooled in their on state while applying a stress voltage.
Under the current collapse, the width of the Coulomb diamond in the $V_\mathrm{g}$ axis direction increases compared to standard measurements.
From the Coulomb diamond, we evaluate the electrostatic capacitance between quantum dots and each electrode, which can be explained by electron trapping by surface states.
The gate length dependence of Coloumb diamond measurement shows that the current collapse can form an effective short gate length, which is the potential method for tuning the quantum dot formation that is difficult to control artifactually in a simple FET structure. 
Our findings suggest that the quantum dots in GaN FETs can serve as a probe to the current collapse, and external control of quantum dot formation in simple FETs is enabled by utilizing the current collapse, conversely.

\section{Acknowledgements}

We thank S. Yamaguchi, 
RIEC Fundamental Technology Center and the Laboratory for Nanoelectronics and Spintronics for writing and technical support. 
Part of this work is supported by 
Rohm Collaboration Project, 
MEXT Leading Initiative for Excellent Young Researchers, 
Grants-in-Aid for Scientific Research (21K18592, 23H01789, 23H04490), 
Iketani Science and Technology Foundation Research Grant,
and FRiD Tohoku University.

\section{AUTHOR DECLARATIONS}
\subsection{Conflict of Interest}
The authors have no conflicts to disclose.
\subsection{Author Contributions}
\textbf{Takaya Abe:} Conceptualization (lead); Data Curation (lead); Formal Analysis (lead); Investigation (lead); Methodology (lead); Visualization (lead); Writing/Review \& Editing (equal). 
\textbf{Motoya Shinozaki:} Conceptualization (equal); Data Curation (equal); Formal Analysis (equal); Investigation (equal); Methodology (equal); Visualization (equal); Writing/Original Draft (lead); Writing/Review \& Editing (equal);
\textbf{Kazuma Matsumura:} Investigation (equal); Methodology (equal); Resources (equal); Writing/Review \& Editing (equal);
\textbf{Takumi Aizawa:} Investigation (equal); Methodology (equal); Writing/Review \& Editing (equal);
\textbf{Takeshi Kumasaka:} Methodology (equal); Resources (equal); Writing/Review \& Editing (equal);
\textbf{Norikazu Ito:} Resources (equal); Writing/Review \& Editing (equal);
\textbf{Taketoshi Tanaka:} Resources (equal); Writing/Review \& Editing (equal);
\textbf{Ken Nakahara:} Resources (equal); Writing/Review \& Editing (equal);
\textbf{Tomohiro Otsuka:} Conceptualization (equal); Methodology (equal); Funding Acquisition (lead); Supervision (lead); Writing/Review \& Editing (lead). 
\bibliography{reference.bib}

\end{document}